\documentclass[11pt]{article}

\let\ssection=\section
\renewcommand{\section}{\setcounter{equation}{0}\ssection}

\def\smallcirc{{\raise 0.5pt \hbox{$\scriptstyle\circ$}}}
\def\parag{\hfil\break} 
\def\kikezd{\parag\underbar}
\def\p{{\partial}}
\def\const{{\rm const}}
\def\vx{{\vec x}}
\def\cP{{\cal P}}
\def\cD{{\cal D}}

\def\IR{{\bf R}}
\def\IZ{{\bf Z}}
\newcommand\half{{\scriptstyle{\frac{1}{2}}}}
\usepackage[pdftex]{graphicx}

\begin{document}

\setlength{\baselineskip}{16pt}

\title{The Maslov correction in the semiclassical
Feynman integral}

\author{
P.~A.~Horv\'athy\footnote{e-mail: horvathy@lmpt.univ-tours.fr}
\\
Laboratoire de Math\'ematiques et de Physique Th\'eorique\\
Universit\'e de Tours\\
Parc de Grandmont\\
F-37 200 TOURS (France)
}

\date{\today}

\maketitle

\begin{abstract}
The Maslov correction to the wave function is the jump of $-\pi/2$ 
in the phase when the system passes through a caustic
point.  This phenomenon is related to the second variation
and to the  geometry of paths, as conveniently
explained in Feynman's path integral framework. 
The results can be extended to any system using the
semiclassical approximation. The $1$-dimensional 
harmonic oscillator is used to illustrate the
different derivations  reviewed here.
\end{abstract}

arXiv~: \texttt{quant-ph/0702236}

\vspace{-5mm}
\section{Introduction}

A fascinating fact, first observed at the end of the 19th century
 \cite{Gouy}, is that, after  passing through a focal point,
 the phase of light jumps by  $-\pi/2$. If 
 a light beam is split into two parts one of them 
passes through a focal point while the other does not,
then, when the two partial waves are recombined,
 a destructive interference is observed.
 
This curious phenomenon has later been extended to the quantum mechanics
of massive particles,
 where it is referred to as the ``Maslov correction''  \cite{Keller,Maslov,Arnold}.
A simple illustration  is provided by the harmonic
oscillator \cite{JMS}. 
Feynman's path integral framework 
\cite{FeynmanHibbs}
is ideally suited to understand how this comes about \cite{HPA}. 

The aim of this Review  is to  derive the  Maslov  correction and  study some if its
aspects  in the path integral framework.

\section{The Feynman propagator of the oscillator}\label{Feynmanprop}

Let us first consider, for simplicity, a problem in $1$ space
dimension. Let 
 $x_1$ and $x_2$ two points and let $T>0$ a time interval \footnote{It is tacitly assumed that the system is conservative. Were this not the case, one
 should consider two instants $t_1$ and $t_2>t_1$ instead of 
 $T=t_2-t_1$.}.
 Let us indeed consider the set, denoted by   ${\cal P}$
of all  curves  $\gamma(t)$ between $x_1$ and $x_2$, i. e. 
 such that $\gamma(0)=x_1$ and $\gamma(T)=x_2$.
 The classical action is a real valued function defined on  ${\cal P}$,
\begin{equation}
S(\gamma)=\int_0^T\!\! L\big(\gamma(t),\dot{\gamma}(t)\big)dt,
\label{hatas}
\end{equation}
where $L(x,\dot{x})$ is the Lagrange function of the system. 
Then the Principle of
Least Action (Hamilton's Principle) tells us that the actual 
motion
which starts in $x_1$  and  arrives to  $x_{2}$ in $T$
 is the  $\bar{\gamma}(t)$ that  makes  $S$  stationary.

Apart of exceptional cases, (see below) 
 the two given points $x_1$ to $x_2$ are joined,  in time $T$, by a unique classical motion curve  $\bar\gamma(t)$.
But in Quantum Mechanics the situation is different. Intuitively, 
a quantum particle moves
not along a unique path, but along {\it all paths} which join these points
in the given time~! Feynman,
in his Thesis, argued indeed that to any  path $\gamma$ in $\cP$ is associated a complex number, namely
\begin{equation}
\exp\left[\frac{i}{\hbar}S(\gamma)\right],
\label{Feynmanfactor}
\end{equation}
where $S$ denotes the classical action  (\ref{hatas})
calculated along  $\gamma$. Next, 
if the ``amplitude'' (whose square $|\psi|^2$ is the probability)
of finding our particle in the instant
$t=0$ at the point $x_{1}$ is $\psi(x_{1})$, then the amplitude, $\psi_{T}(x_{2})$,
of finding it at $x_{2}$ at time $T$ will be
\begin{equation}
    U_T\psi(x_2)\equiv\psi_T(x_2)=\int_{-\infty}^{\infty}
K(x_2,T|x_1,0)\psi(x_{1})dx_{1}.
\label{idofej}
\end{equation}
Here the propagator, $K(x_2,T|x_1,0)$, which describes the transition from
$x_{1}$-b\H ol $x_{2}$ in time $T$, is, says Feynman, a ``sum'' of the  
contributions (\ref{Feynmanfactor}) of all paths, 
\begin{equation}
K(x_2,T|x_1,0)=\int_\cP\exp\left[\frac{i}{\hbar}S(\gamma)\right]
\cD\gamma.
\label{propag}
\end{equation}

But all this intuitive, unless we say what
 ``$\cD\gamma$'' means here.
The definition of the integration measure is indeed the main stumbling block of the
whole theory, and, despite many efforts, no fully satisfactory 
answer is available as yet.
The miracle is that the integral can, in some cases, nevertheless evaluated \cite{FeynmanHibbs}.
Below we present one possible method we illustrate
 on the example of a {\it one dimensional harmonic oscillator}.


Let us indeed consider an arbitrary path $\gamma(t)$ in $\cP$ that satisfies the
given boundary conditions. Let us assume that $\cP$ contains a unique
classical path, $\bar{\gamma}(t)$, 
 and let us decompose  $\gamma$ into the sum of $\bar{\gamma}(t)$ and a
\begin{equation}
\gamma(t)=\bar{\gamma}(t)+\eta(t),
\qquad\hbox{ahol}\qquad
\eta(0)=\eta(T)=0,
\label{eta}
\end{equation}
since the end points are kept fixed. The Lagrange function of the oscillator 
is 
$
L=\frac{m}{2}\big({\dot{x}}^2-\omega^2x^2\big). 
$
 The action along $\gamma(t)$ is therefore
\begin{equation}
S(\gamma)=\int_0^T\!\half m\big(\dot{\bar{\gamma}}^2
-\omega^2{\bar{\gamma}}^2\big)dt
+
m\int_0^T\!\big(\dot{\bar{\gamma}}\,\dot{\eta}
-\omega^2\bar{\gamma}\,\eta\big) dt
+
\half m\int_0^T\!(\dot{\eta}^2
-\omega^2{\eta}^2)dt.
\label{hatkifejtes}
\end{equation}
For the oscillator,
$
\ddot{x}+\omega^2x=0
$, so that
$
x(t)=A\sin\omega t +B\cos\omega t,
$
where the constants $A$ and $B$ are determined by the initial conditions,
$$
B=x_1\qquad\hbox{and}\qquad
A=\frac{x_2-x_1\cos\omega T}{\sin\omega T},
$$
provided $\sin\omega T\neq 0$,i.e., if $\omega T\neq N\pi$.  
Hence
{\it If  $T$ is not integer multiple of the half-period,
\begin{equation}
 T\neq N\times\frac{\tau}{2},
 \qquad\tau=\frac{2\pi}{\omega},
 \label{nemkauszt}
\end{equation}
there exists a unique classical path that starts in $x_1$ and
arrives, after time $T$, to $x_2$-be}.
The  action calculated along this path is
\begin{equation}
S(\bar{\gamma})=\frac{m\omega}{2\sin\omega T}\left(
(x_1^2+x_2^2)\cos\omega T-2x_1x_2\right).
\label{oszchat}
\end{equation}

Returning to (\ref{hatkifejtes}), the first term is
 $S(\bar{\gamma})$. 
Next, integration by parts of the middle term
yields
$$
\int_0^T\!\big(\dot{\bar{\gamma}}\,\dot{\eta}
-\omega^2\bar{\gamma}\,\eta\big) dt=-\int_0^T\!\big(\ddot{\bar{\gamma}}
+\omega^2\bar{\gamma}\big)\,\eta\, dt=0,
$$
since $\bar{\gamma}$ satisfies the classical equation of motion
$\ddot{x}=-\omega^2x$. This term vanishes therefore, providing us with the propagator
\begin{equation}
K(x_2,T|x_1,0)=
\exp\left[\frac{i}{\hbar}S(\bar{\gamma})\right]\times F(T),
\label{oscpropag}
\end{equation}
where the ``reduced propagator'', $F(T)$, is a path integral taken
over all variations,
\begin{equation}
F(T)=\int \exp\left\{\frac{im}{2\hbar}\int_0^T\!\left[\big(
\dot{\eta}^2-\omega^2\eta^2\right]dt\right\}\cD\eta.
\label{F(T)}
\end{equation}

But all this is still intuitive, as we still not say what 
``$\cD\eta$'' actually means.
To  answer this question, Feynman expands the (periodic) variation into a
Fourier series,
\begin{equation}
\eta(t)=\sum_{k=1}^\infty a_k\sin [k\frac{\pi}{T}t]
\quad\Rightarrow\quad
\int_0^T\!{\eta}^2dt=\frac{T}{2}\sum_ka_k^2,
\qquad 
\int_0^T\!\dot{\eta}^2dt
 =\frac{T}{2}\sum_k\big(\frac{k\pi}{T}\big)^2a_k^2.
 \label{four}
\end{equation}
Then he argues as follows. Any path is determined by its Fourier coefficients
$a_{k}$; instead of ``integrating over all paths''
let us integrate over all Fourier coefficients,
\begin{eqnarray}
F(T)=\lim_{n\to\infty}\,{\cal J}\int_{-\infty}^{\infty}\!
\dots \int_{-\infty}^{\infty} \!
\exp\left\{\sum_{k=1}^ni\,\frac{\lambda_k}{2\hbar}a_k^2
\right\}\times da_1\dots da_n
\label{sokintegral}
\end{eqnarray}
where\vspace{-5mm}
\begin{eqnarray}
\lambda_k=m\left(
\big(\frac{\pi k}{T})^2-\omega^2\right)
\label{eigenvalue}
\end{eqnarray}
and  where ${\cal J}$ denotes the Jacobian of the (linear)
transformation  $\cP\to \Big\{\hbox{Fourier coefficients}\Big\}$.
Let us observe that ${\cal J}$ is independent of the data ($\omega, m$, etc)
of the oscillator and even of $\hbar$.

 The classical Fresnel integrals can be evaluated,
\begin{equation}
\int_{-\infty}^{\infty}\exp\left[i\frac{\lambda}{2}x^2\right]dx=
\sqrt{\frac{2\pi}{\lambda}}\,e^{i\frac{\pi}{4}}.
\label{efresnel}
\end{equation}
Then the sum in the exponent in (\ref{sokintegral})  becomes a
product, 
\begin{equation}
F(T)=\lim_{n\to\infty} C_{n}
\left(\prod_{k=1}^n\lambda_k\right)^{-1/2},
\label{faktorok}
\end{equation}
where  $C_{n}$ is the product of various (divergent) factors.

The product of eigenvectors can be split into two parts.
\begin{equation}
\prod_{k=1}^n\lambda_k=
\prod_{k=1}^nm\frac{k^2\pi^2}{T^2}\times
\prod_{k=1}^n\Big(1-\frac{\omega^2T^2}{k^2\pi^2}\Big).
\label{tenyezok}
\end{equation}
The Euler formula says now that
\begin{equation}
\prod_{k=1}^\infty\left(1-\frac{x^2}{k^2\pi^2}\right)=
\frac{\sin x}{x},
\label{Euler}
\end{equation}
so that
$$
F(T)=C\sqrt{\frac{\omega T}{\sin\omega T}},
$$
where $C$ denotes the product of all $\omega$-independent factors.

Let us now remember that for $\omega\to 0$ we get a free particle,
and therefore 
\begin{equation}
F^{free}(T)=\sqrt{\frac{m}{2\pi i\hbar T}}.
\label{szabadredprop}
\end{equation}
This yields  $C$, and inserting the action (\ref{oszchat}) the propagator becomes, at last,
\begin{equation}
K(x_2,T|x_1,0)=\left(\frac{m\omega}{2\pi i\hbar\sin\omega T}\right)^{1/2}\!
\times\exp\left\{\frac{im\omega}{2\hbar\sin\omega T}
\big[(x_1^2+x_2^2)\cos\omega T-2x_1x_2\big]\right\}.
\label{Fprop}
\end{equation}

\section{The phase correction}
 

Let us observe that
 if the ``$\sin$''   in the denominator  vanishes,
the whole expression becomes meaningless. How should we then
continue after such a singular point.
The answer comes from the evaluation of the Fresnel integrals~:
 Euler's formula, (\ref{efresnel}), is only valid for $\lambda>0$.
But
\begin{equation}
\lambda_k>0\qquad \Longleftrightarrow\qquad
0<T<k\frac{\tau}{2},
\qquad
\end{equation}
where  $\tau=2\pi/\omega$ is the period. Before reaching
the first half period,
$0<T<\tau/2$, all factors under the square root are positive and
Feynman's calculation is correct. After the first half period
(but before a full periode), however, i. e. for $\tau/2<T<\tau$,
the first factor under the root is negative, while all 
the other factors remain positive.
$$
\lambda_1<0,\qquad \lambda_k>0, \quad k\geq2.
$$ 
As a result, the propagator gets multiplied by
\begin{equation}
\frac{1}{\sqrt{-1}}=-i=e^{-i\frac{\pi}{2}}.
\end{equation}
Thus, the {\it phase of the propagator (and therefore also
of the wave function) jumps by $(-\pi/2)$.}

Similarly, after $N$  but before  $(N+1)$, half-periods, i.e. for
\begin{equation}
N\times\frac{\tau}{2}<T<(N+1)\times\frac{\tau}{2},
\label{N}
\end{equation}
the first $N$ factors in (\ref{faktorok}) become negative. The 
phase jumps, therefore, by $N\times(-\pi/2)$.
The correct result is hence 
\begin{equation}
    \begin{array}{ll}
K(x_2,T|x_1,0)=
&\left(\displaystyle{\frac{m\omega}{2\pi\hbar|\sin\omega T|}}\right)^{1/2}
\!e^{-\frac{i\pi}{4}} 
\\[16pt]
&\times\exp\left\{\displaystyle{\frac{im\omega}{2\hbar|\sin\omega T|}}
\times[(x_1^2+x_2^2)\cos\omega T-2x_1x_2]
\right\}\times e^{-\frac{i\pi}{2}N}.
\end{array}
\label{Masprop}
\end{equation}

We must admit that our argument has been somewhat sloopy: each
factor
$\sqrt{-1}$ could be $i$ or $-i$, and choosing the second one would
change every phase jump
from $-\pi/2$ to  $\pi/2$. Worse: every jump can be chosen independently~!
Which square root of $(-1)$ has to be chosen~?
During the first half-period, the question is irrelevant, since it is
merely a global phase. But repeating it $N$ times it {\it is} relevant~!

The formula that generalizes (\ref{efresnel}) can be derived through
analytic extension.
For any real $\lambda\neq0$,
\begin{equation}
\int_{-\infty}^{\infty}\exp\left[i\frac{\lambda}{2}x^2\right]dx
=\left\{
\begin{array}{cc}
\displaystyle(\frac{2\pi}{\lambda})^{1/2}e^{i\frac{\pi}{4}}
&\lambda >0
\\[14pt]
\displaystyle(\frac{2\pi}{-\lambda})^{1/2}e^{-i\frac{\pi}{4}}
&\lambda <0
\end{array}\right..
\label{Fresnel}
\end{equation}
Euler's formula, (\ref{Euler}), only holds for
$x<\pi$, and  otherwise it should be replaced by
\begin{equation}
\prod_{k=1}^\infty\left|1-\frac{x^2}{k^2\pi^2}\right|=
\frac{|\sin x|}{x}. 
\qquad x>0,
\label{Eulerbis}
\end{equation}

This confirms the validity of our previous argument~: passing through
every half-peridod
contributes a new negative $\lambda$, and this changes
the phase by  $\pi/2$.

What happens for  $T=N\times\tau/2$~?
The propagator plainly diverges, since
$\sin\omega T\to \sin N\pi=0$ in the denominator.
To derive the correct result, let us use the (semi-)group property
of the time evolution,
\begin{equation}
U_{t+t'}=U_t\smallcirc U_{t'}
\qquad\Rightarrow\qquad
U_{N\frac{\tau}{2}}=(U_{\frac{\tau}{4}})^{2N}.
\end{equation}
According to (\ref{Fprop})
\begin{equation}
U_{\frac{\tau}{4}}\psi(x_2)=(\frac{m\omega}{2\pi\hbar})^{1/2}
e^{-i\pi/4}\times
\int e^{-i\frac{m\omega}{\hbar}x_1x_2}\psi(x_1)dx_1,
\end{equation}
which is, essentially, a Fourier-transformation. 
But the square of such  a Fourier-transformation carries any function into
itself up to a change of its argument. Hence
\begin{equation}
\psi_{N\frac{\tau}{2}}(x_2)=e^{-i\frac{\pi}{2} N}\psi((-1)^Nx_2),
\end{equation}
i.e.
\begin{equation}
K(x_{2},T|x_{1},0)=\exp\left[-i\frac{\pi}{2}N\right]\times
    \delta\big(x_{1}-(-1)^Nx_{2}\big).
    \label{kausztprop}
\end{equation}

Going to $D$ dimensions, the phase will jump by 
$D\times\pi/2$. What happens if the oscillator is not perfectly harmonic~?
One has to study higher-order terms \cite{Schulman}.

The result can be extended to several similar situations.

$\bullet$ 
For a {\it forced oscillator} driven by a constant external force \cite{FeynmanHibbs,Cheng,LiangMorandi}, 
 the previous calculation can be repeated
 word-by-word. The Lagrange function is
\begin{equation}
L=\frac{m}{2}\big({\dot{x}}^2-\omega^2x^2\big)+fx, 
\label{forcedoszc}
\end{equation}
where $f=\const$. For $T\neq k\tau/2$ the Hamiltonian action is 
\begin{equation}
\begin{array}{ll}
S_f(\bar{\gamma})=&\displaystyle\frac{m\omega}{2\sin\omega T}\left(
(x_1^2+x_2^2)\cos\omega T-2x_1x_2\right)
\\[14pt]
&+2f\displaystyle\frac{(1-\cos\omega T)}{m\omega^2}(x_1+x_2)
-f^2\displaystyle\frac{2(1-\cos\omega T)-\omega T\sin\omega T}{m^2\omega^4}
\end{array}
\label{focedoszchat}
\end{equation}
cf. (\ref{oszchat}). The propagator is again
 (\ref{Masprop}) with the only change that the extra terms
in  (\ref{focedoszchat}) should be accounted for. In fact 
\cite{LiangMorandi},
\begin{equation}
K_f(x_2,T|x_1,0)=\exp\left[i\displaystyle\frac{f^2T}{2m\omega^2\hbar }\right]K_{osc}(x_2-x^*,T|x_1-x^*,0)
\end{equation}
where $x^*=f/m\omega^2$. This latter formula also holds for
$T=N\times\tau/2$.


\section{Derivation from the wave function}

Before further investigating the various aspects and derivations 
of the Maslov correction from the
Feynman integral viewpoint, we would like to show how it can be understood using
our  knowledge of the solution of the Schr\"odinger
equation.
The clue \cite{LiangMorandi} is to write the propagator as
\begin{equation}
K(x_1,x_2;T)=\sum_{n=0}^\infty\exp[-i(n+\half)\omega T]\psi_n(x_1)\psi_nx_2,
\label{propwf}
\end{equation}
where the $\psi_n$ s are the normalized harmonic oscillator wavefunctions~:
\begin{equation}
\psi_n(x)=\left(\frac{1}{2^nn!}\sqrt{\frac{m\omega}{\pi\hbar}}\right)^{1/2}
\,\exp(-\half\zeta^2)H_n(\zeta),
\label{owf}
\end{equation}
where $\zeta=\sqrt{m\omega/\hbar}x$ and the $H_n$ denote the Hermite
polynomials.
Their  property important for us is their behavior under space reflection,
$
\psi_n(-x)=(-1)^n\psi_n(x),
$ 
implied by that of the Hermite polynomials.

(\ref{propwf}) can now be evaluated. 
Let us assume that we are not in caustic, $t\neq N\pi/\omega$, where $N$ is some integer.
Then  write
$ 
\omega t=N\pi+\omega\tau,
$
$0<\tau<\pi/\omega$.
$N$ is hence the integer part of $\omega t/\pi$. Now from (\ref{propwf}), we infer that
\begin{equation}
K=\exp(-\half iN\pi)\,\sum_{n=0}^\infty\exp[-inN\pi-(i(n+\half)\omega\tau]\psi_n(x_1)\psi_n(x_2).
\end{equation}

\begin{equation}
K(x_1,x_2;T)=\exp(-\half i\pi{N})\sum_{n=0}^\infty
\exp[-i(n+\half)\omega t]\psi_n(x_1)\psi_n((-1)^Nx_2).
\label{propwfallt}
\end{equation}

The  Hermite polynomials admit a generating function,  namely
\begin{equation}
{\cal G}(\zeta_1,\zeta_2;z)=
\frac{1}{\sqrt{1-z^2}}\exp\left(\frac{2z\zeta_1\zeta_2-z^2(\zeta_1^2+\zeta_2^2)}{1-z^2}\right).
\label{Hgenfunc}
\end{equation}
${\cal G}$ is analytic over the $z$ plane  with cuts going from
$-\infty$ to $-1$ and from $+1$ to $+\infty$.
For $|z|<1$ the generating function admits the expansion
\begin{eqnarray}
{\cal G}(\zeta_1,\zeta_2;z)=\sum_{n=0}^\infty\frac{z^n}{n!2^n}\,H_n(\zeta_1)H_n(\zeta_2).
\label{Hgenfuncexp}
\end{eqnarray}
The product of the Hermite polynomials is expressed therefore, as
\begin{eqnarray}
H_n(\zeta_1)H_N(\zeta_2)=2^n\frac{\ d^n}{dz^n}{\cal G}(\zeta_1,\zeta_2;z)\bigg|_{z=0}.
\end{eqnarray}
Then putting $z=\exp(-\alpha i),\, 0<\alpha<\pi$, a lengthy calculation
\cite{LiangMorandi}  yields
\begin{equation}
{\cal G}=\exp[\half(i\alpha+\zeta_1^2+(\zeta_2)^2)]\frac{e^{-i\pi/4}}{2\sin\alpha}
\exp\left(\frac{i}{2\sin\alpha}[\zeta_1^2+(\zeta_2)^2\cos\alpha-2\zeta_1\zeta_2]\right).
\end{equation}
Collecting our formulae, 
\begin{equation}
K(x_1,x_2,T)=\sqrt{m\omega/\pi\hbar}\exp(-\half\pi{N}[\frac{\omega t}{\pi}]
\exp[-\half(i\omega\tau+\zeta_1^2+\zeta_2^2)]{\cal G}(\zeta,(-1)^k\zeta_2;e^{-i\omega\tau}).
\end{equation}
Noting finally that
$
\sin(\omega\tau)=|\sin(\omega\tau)|=(-1)^k\sin(\omega t)
$ and
$
\cos(\omega\tau)=(-1)^k\cos(\omega t)
$
we get precisely the propagator (\ref{Fprop}), valid for times which are not integer multiples of half-periods.

In a caustic i.e. if $t=(\pi/\omega)N$,
then, letting $\omega t\to N\pi$ in (\ref{propwfallt}) yields \cite{LiangMorandi}, using the completeness of the wave functions,
the formula valid in the caustic point, i.e., 
\begin{equation}
\lim_{\omega t\to N\pi}K(x_1,x_2;t)=
\exp(-\half iN\pi)\delta(x_1-(-1)^Nx_2)
\end{equation}
cf. (\ref{kausztprop}).

Let us mention that the Maslov correction can also be derived from a semiclassical analysis of the wave function \cite{Keller}.

\section{How minimal is the ``minimal action''~?}

Let us now return to classical mechanics. The classical hamiltonian action
is, as explained in Section \ref{Feynmanprop}, a real function, $S(\gamma)$,
defined on the ``infinite dimensional manifold'',  $\cP$,
of all paths which join $x_{1}$ to $x_{2}$ in times $T$. 
A variation $\eta$  can be viewed  in turn as
a ``tangent vector'' to  $\cP$ in $\gamma$ \cite{HPAU}, cf. Fig. 1.
\begin{figure} 
\begin{center}
\includegraphics[scale=.60]{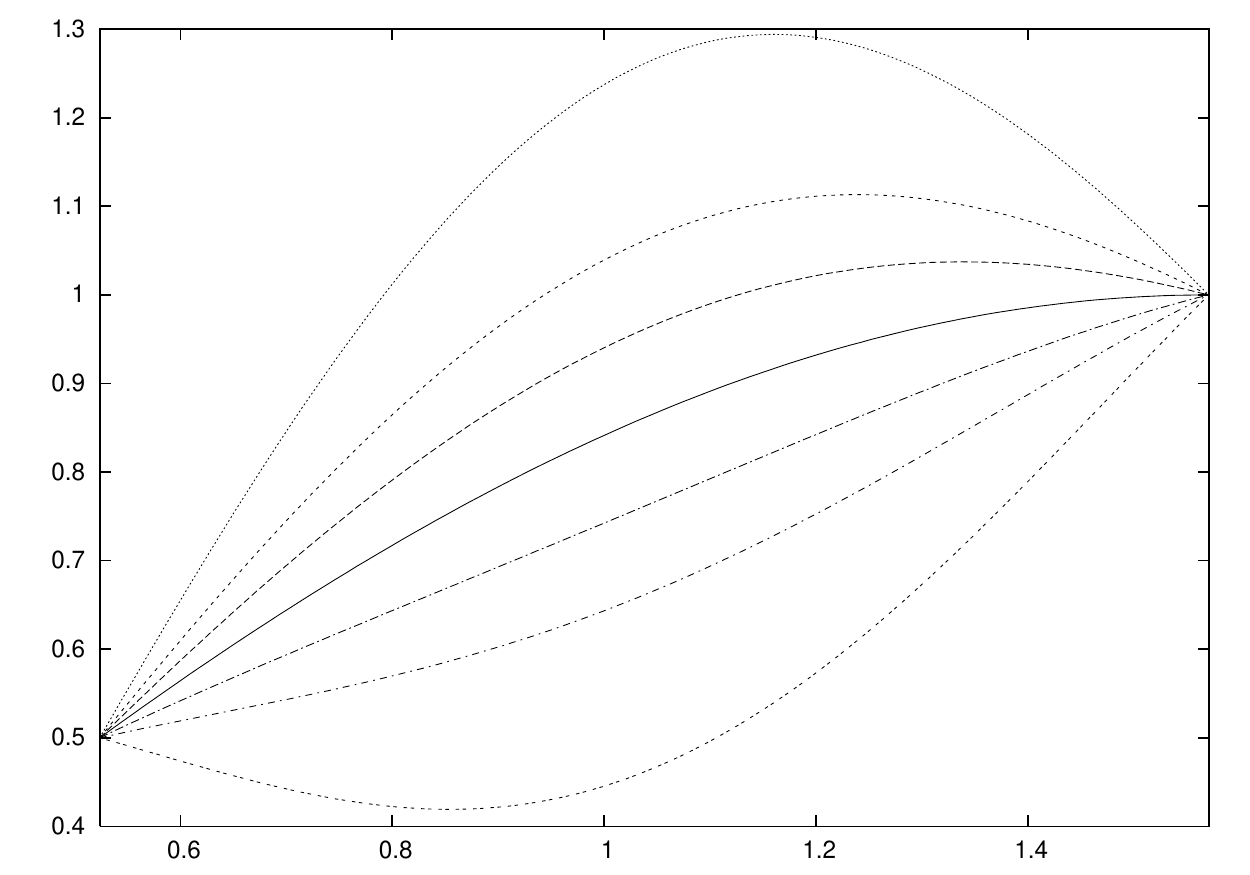}
\label{varcurves.pdf}
\caption{Those paths which join $x_1$ to $x_2$ in time $T$ form an
``infinite dimensional manifold'' $\cP$. The the actual motion, 
$\bar\gamma(t)$, is
a critical point of the hamiltonian action, viewed as a real-valued
function on $\cP$.}
\end{center} 
\end{figure}
$\delta S$, the
first variation of the action, is the directional derivative of
$S$ in the ``point'' $\gamma$,
\begin{equation}
    \delta S_{\gamma}(\eta)=\lim_{s\to0}
    \frac{S(\gamma+s\eta)-S(\gamma)}{s}.
  \label{1vari}
\end{equation}
$\delta S_{\gamma}$ is, hence a one-form on  $\cP$.
Just like in finite dimensional calculus, if
$\bar{\gamma}$ is an extremal point of $S$, then  
the directional derivative must vanish in any direction,
\begin{equation}
    \delta S_{\bar{\gamma}}\equiv\delta S_{\bar{\gamma}}(\eta)=
    \int\left\{\frac{\ d}{dt}
    \Big(\frac{\p L}{\p \dot{x}}\Big)-\frac{\p L}{\p x}\right\}\eta dt=0.
    \label{variegy}
\end{equation}
This yields the classical (Euler-Lagrange) equations,
$
\frac{\ d}{dt}
    \Big(\frac{\p L}{\p \dot{x}}\Big)-\frac{\p L}{\p x}=0,
$
whose solution satisfying the boundary conditions, $\bar{\gamma}$ is, by Hamilton's Principle, is the actual motion.

But does such a solution always exist, and if it does, is it unique~?
The answer is, generically, affirmative. Not always, however. This 
point can  again illustrated by the oscillator. Let us indeed
assume that the elapsed time is an integer multiple of the half-period,
\begin{equation}
T= N\times\frac{\tau}{2}.
\label{kauszt}
\end{equation}
Then, independently of the initial velocity, {\it all} motions which start in $x_1$ arrive, at time $T$, to $x_2=(-1)^Nx_1$.
(cf. Fig. 2.).
\goodbreak 
\begin{figure}
\begin{center}
\includegraphics[scale=.60]{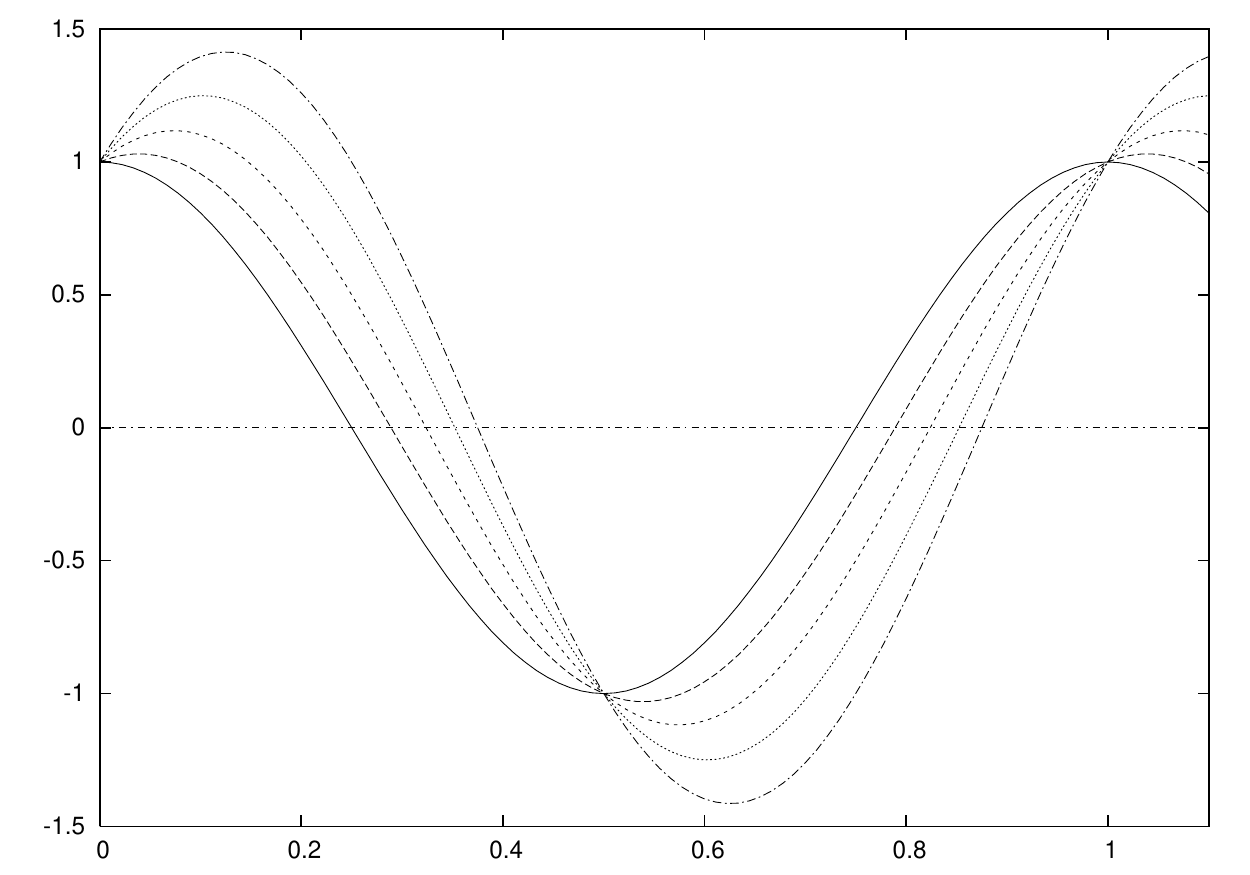}
\end{center} 
\label{kausztika}
\caption{After a half period, 
all trajectories meat again,
independently
of the initial velocity,  in the same point opposite to the one they started from.}
\end{figure}
For this particular time,  (i) either i.e. when 
$x_2\neq(-1)^Nx_1$, there is no classical path at all with the required boundary condition, (ii) or i.e. for $x_2=(-1)^Nx_1$, there are infinitely many of them.

The solutions of the classical equation of motion behave, hence, as
light rays that start from one focus of an elliptic mirror: after reflection,
they are all collected into the same point, namely the other focus.
(By analogy, such point are also called focal points also in variational mechanics).

Apart of this particular situation, there is a unique classical path,
$\bar{\gamma}$, in $\cP$.
Let us assume that we are in this, generic, situation. Is the action minimal~?
Let us emphasized that, just like in finite dimension, (\ref{variegy}) 
is merely a {\it necessary} condition for having a minimum. It is not
sufficient, though. In other words, $\bar{\gamma}$ is a critical point of
 $S$, but not necessarily a minimal one.
 
Let us mention, at this point, that the variational calculus is in fact local,
i.e., it is a differential calculus in the neighborhood of a path $\bar\gamma$.
When calculating the first variation, we compare in fact the value of
$S$ calculated along paths 
which can be smoothly deformed to $\bar\gamma$. 
All such paths belong to
a single path component of $\cP$. If the latter has more then one
path components, as it happens in the,
one has to perform the variational calculus in each path-connected
sector separately.

In technical terms, a ``point'' (i.e. a path in a manifold $M$)
 $\gamma$ belongs to the
path-component of  $\bar\gamma$ if
$\gamma$  can joined to the ``point''  $\bar\gamma$ in  $\cP$
if there exists a continuous ``path'' of ``points'' $\gamma_s$ 
such that 
$ 
\gamma_1=\gamma,
\,
\gamma_0=\bar\gamma.
$
But this means precisely that $\bar\gamma$
can be deformed into  $\bar\gamma$ by a homotopy.
The path components of $\cP$ are therefore labeled with the
homotopy classes of the underlying space,
$
\pi_0(\cP)=\pi_1(M).
$
 
An illustration is provided by the  Aharonov-Bohm experiment  \cite{AB},
where $M$ is the punctured plane $\IR^2\setminus\{0\}$,
whose $\pi_1$ is $\IZ$. The (free) action has, hence, 
a minimum in two homotopy classes, namely in those which pass near
the  solenoid (assumed infinitely thin) on either side. 
In all other path components, i. e. for paths which turn around the
origin, the action has no critical point.

Returning to the critical points of the action, they behave just like
those of a function defined over a finite ($D$) dimensional space~:
$\vx_0$ is a critical point of a function $V(\vx)$ if
$$
\delta V(\vx_0)=0
\qquad\hbox{i.e.}\qquad
\p_iV(\vx_0)=0\qquad\forall \ i.
$$
Whether a given critical point $\vx_0$ is a  
minimum, a maximum or a saddle point,  can be determined by looking at
the matrix formed from second-order partial derivatives,
\begin{equation}
\delta^2 V\equiv \delta^2 V(\vx_0)=
\frac{\p^2V}{\p x^i\p x^j}(\vx_0).
\end{equation}
Is the matrix of  $\delta^2 V$ definite~? Being symmetric, $\delta^2 V$
will have $D$ (real) eigenvalues, 
$$
\delta^2 V\,e_a=\lambda_a\,e_a,
\qquad
a=1,\dots, D.
$$
Then $\vx_0$ is
\begin{equation}
\left\{\begin{array}{lll}
\hbox{a minimum if}\;&\lambda_a>0\,&\forall a
\\[8pt]
\hbox{a maximum if}\;&\lambda_a<0\,&\forall a
\\[8pt]
\hbox{saddle point if}\qquad &\lambda_a>0\quad\lambda_b<0\;&\hbox{for some}\; a, b.
\end{array}\right.
\end{equation}

In an analogous way, to see if a given classical path $\bar\gamma$ makes
the action  minimal one or not, we have to calculate the second variation,
$\delta^2S_{\bar{\gamma}}$.
This is a quadratic form on the ``tangent space'' of the ``variations''.
To be a minimum, the second variation has to be positive definite,
\begin{equation}
    \delta^2S_{\bar{\gamma}}(\eta,\eta)>0
    \qquad\forall \eta.
    \label{minfelt}
\end{equation}
If, however,
\begin{equation}
    \delta^2S_{\bar{\gamma}}(\eta,\eta)<0
    \qquad\hbox{but}\qquad
    \delta^2S_{\bar{\gamma}}(\eta',\eta')>0
    \label{nyeregfelt}
\end{equation}
for suitable variations $\eta$ and $\eta'$, then $\bar{\gamma}$  is a saddle point.

A variation $\eta$ such that
\begin{equation}
\delta^2S_{\gamma}(\eta,\eta)<0
\end{equation}
is a negative mode, and if
\begin{equation}
\half\delta^2S(\eta,\eta')=0
\qquad\forall\eta',
\label{nullter}
\end{equation}
it is a zero-mode. In the direction of a zero mode the function is, in
the first approximation, invariant. In this case, the nature of the
critical point depends on the higher variations.

Let us examine the second variation in some detail.
$$
\half\delta^2S\equiv\half\delta^2S_{\bar{\gamma}}(\eta,\eta)
=\half\int_{0}^T\!\left\{
\frac{\p^2L}{\p x^2}\eta^2+
2\frac{\p^2L}{\p x\p \dot{x}}\eta\dot{\eta}+
\frac{\p^2L}{\p x^2}\eta^2+
\frac{\p^2L}{\p \dot{x}^2}{\dot{\eta}^2}
\right\}dt,
$$
where the integration is along the (assumed unique) classical path $\bar\gamma$.
After partial integration,
\begin{eqnarray}
&\half\delta^2S=\int_{0}^T\!\big(\eta,\Lambda\eta\big)dt,\nonumber
\\[10pt]
&\Lambda=
    -\displaystyle\frac{\ d}{dt}\left(
    \displaystyle\frac{\p^2L}{\p \dot{x}^2}\displaystyle\frac{\ d}{dt}+
    \displaystyle\frac{\p^2L}{\p x\p \dot{x}}\right)
    +
    \left(\displaystyle\frac{\p^2L}{\p x\p \dot{x}}
    \displaystyle\frac{\ d}{dt}+\displaystyle\frac{\p^2L}{\p x^2}
    \right).
    \label{Lambda}
\end{eqnarray}
where $\Lambda$ is the operator of the second variation. $\half\delta^2S$
is hence positive definite if all eigenvalues of the quadratic form
$\Lambda$  are positive, $\lambda>0$, for all
\begin{equation}
    \Lambda\,\eta=\lambda\eta,
    \qquad
    \eta(0)=\eta(T)=0.
    \label{sajert}
\end{equation}

$\bullet$  In the simplest possible case of a one-dimensional massive particle in a potential,
\begin{equation}
    L=\frac{m{\dot{x}}^2}{2}-V(x)
    \qquad\Rightarrow\qquad
    \Lambda=-m\left(\frac{\ d^2}{dt^2}+\frac{d^2V}{dx^2}\right).
    \label{Lambdabis}
\end{equation}   
For the $1D$ oscillator, e.g.,
\begin{equation}
    \Lambda_{osc}=-m\left(\frac{\ d^2}{dt^2}+\omega^2\right).
    \label{oscLambda}
\end{equation}
Hence, taking into account the boundary conditions,
\begin{eqnarray*}
m(\ddot{\eta}+\omega^2\eta\big)=-\lambda\eta
\quad\Rightarrow\quad
\eta(t)=\sin[\big(\omega^2+\frac{\lambda}{m}\big)^{1/2}\,t].
\end{eqnarray*}
But $T\sqrt{\omega^2+\frac{\lambda}{m}}=k\pi,\; k=0,\pm1,\dots$
due to periodicity, so that
\begin{equation}
    \lambda_{k}=m\left(\big(\frac{k\pi}{T}\big)^2-\omega^2\right)
    \qquad\hbox{and}\qquad
    \eta_{k}(t)=\sin(\frac{k\pi}{T}t),
    \label{oscsajert}
\end{equation}
cf. (\ref{four}) and (\ref{eigenvalue}).
The  integer $N>0$ in (\ref{N}) 
$
N\tau/2<T<(N+1)\tau/2,
$
counts hence the negative eigenvalues.
Thus, the oscillator trajectory is a minimum of the action 
during the first half period. For time beyond $N\tau/2$, it becomes
a saddle point with $N$ negative modes~!

What happens in the focal points~? For $T=N\times \tau/2$,
as seen above, either we have no classical path at all or we
have infinitely many between our fixed endpoints. Let $x_2=(-1)^Nx_1$
and let us consider some $\bar{\gamma}$. Then all the other classical
paths can be viewed as a variation of $\bar{\gamma}$, labeled by a parameter
$s$, $\gamma=\gamma_s$. The action calculated for any such classical path
$\gamma=\gamma_s$ is the same. Varying $s$ we get, hence, a ``curve'' in $\cP$
inscribed onto the ``level surface'' $S=\const$. The derivative 
w.r.t. $s$ of $S(\gamma_s)$ is therefore zero, so that 
\begin{equation}
\eta(t)=\frac{\ d}{ds}\gamma_s(t)\bigg|_{s=0}
\end{equation}
is a null mode at $\bar\gamma$.

Minimum or not~: what are the physical consequences~?
At the purely classical level, nothing at all~: (\ref{variegy})
yields the correct equations of motion in all cases.
In Quantum Mechanics,  however, the consequence is precisely the Maslov
phase jump, as we explain it below.  

Let us first consider another example.


$\bullet$ A phase jump similar to the one found for the oscillator is observed 
for a charged particle moving perpendicularly to the induction lines of a
constant magnetic field \cite{FeynmanHibbs}. Classically, the particle rotates uniformly with
 Larmor frequency ${eB}/{m}$ i.e. with 
period $\tau=2\pi m/eB$.
The Lagrange function is
\begin{equation}
L_B=
\frac{m}{2}\left(\dot{x}^2+\dot{y}^2+
2\omega\big(x\dot{y}-y\dot{x}\big)\right),
\qquad
\omega=\frac{eB}{2m}.
\end{equation}

If $T$ is not an integer multiple of a (full) period,
$
T\neq N\times\tau=\frac{2N\pi m}{eB},
$
then there is a unique classical trajectory that links any
two points $(x_1,y_1)$ and $(x_2,y_2)$ in time  $T$. The 
action calculated for it is
\begin{equation}
S_B=\frac{m\omega}{2}
    \Big[\big[(x_{2}-x_{1})^2+(y_{2}-y_{1})^2\big]
    \cot(\omega T)
    +2(x_{1}y_{2}-x_{2}y_{1})\Big].
\label{Bhatas}
\end{equation}
The matrix of the second variation is now
\begin{equation}
    \Lambda_{B}=m
    \left(\begin{array}{cc}
    \displaystyle\frac{\ \ d^2}{dt^2}&-2\omega\displaystyle\frac{\ d}{dt}
    \\[8pt]
    2\omega\displaystyle\frac{\ d}{dt}
    &\displaystyle\frac{\ \ d^2}{dt^2}
    \end{array}
    \right).
    \label{BLambda}
\end{equation}
Our task is to solve  the eigenvalue problem 
\begin{equation}
\left\{\begin{array}{ll}
m\ddot{\eta}_x-2m\omega\dot{\eta}_y=-\lambda\eta_x
\\[8pt]
m\ddot{\eta}_y+2m\omega\dot{\eta}_x=-\lambda\eta_y
\end{array}\right.,
\qquad
\eta_x(0)=\eta_y(0)=0=\eta_x(T)=\eta_y(T),
\label{Beigen}
\end{equation}

The solutions are readily derived by separating the equations (\ref{Beigen}) 
by applying a time-dependent rotation, 
\begin{eqnarray*}
\left\{\begin{array}{ll}
\eta_x=\ \ \cos\omega t\,\xi+\sin\omega t\,\zeta
\\[8pt]
\eta_y=-\sin\omega t\,\xi+\cos\omega t\,\zeta
\end{array}\right.
\quad\Rightarrow\quad
\left\{\begin{array}{ll}
m\ddot{\xi}+(m\omega^2+\lambda)\xi=0
\\[8pt]
m\ddot{\zeta}+(m\omega^2+\lambda)\zeta=0
\end{array}\right..
\end{eqnarray*}
By periodicity,
$
\xi,\zeta\propto\sin\big(\sqrt{\omega^2+\lambda}\,t\big)
=\sin\frac{k\pi}{T}t,
$
where $k$ is some integer. The eigenvalues are, therefore, doubly
degenerate, and are identical to those in the oscillator problem~:
\begin{equation}
    \lambda_{k}=m\left(\big(\frac{k\pi}{T}\big)^2-\omega^2\right),
    \qquad
    k=0,1,\dots.
    \label{Bsajert}
\end{equation}
The {\it reduced propagator is hence identical to that of a
planar oscillator whose frequency is half of 
the Larmor value, }  
$\omega=eB/2m$.

 If $N\tau<T<(N+1)\tau$, then
\begin{equation}\begin{array}{ll}
    K(x_2,y_2,T|x_1,y_1 0)=\displaystyle
    \frac{m\omega}{2\pi i\hbar|\sin\omega T|}\times
    \\[18pt]
    \exp
    \left\{\displaystyle\frac{im\omega}{2\hbar}
    \Big[\big[(x_{2}-x_{1})^2+(y_{2}-y_{1})^2\big]\cot(\omega T)
    +2(x_{1}y_{2}-x_{2}y_{1})\Big]\right\}(-1)^N,
    \quad\omega=\displaystyle\frac{eB}{2m}.
    \end{array}
    \label{Bprop}
\end{equation}
cf. \cite{Cheng2,LiangMorandi}.
After  $N$ full periods i.e. at
$
T=\frac{2N\pi m}{eB}
$
[which corresponds to $N$ half-oscillator-periods],
all classical motions meet again in the point they started from.
According to our previous results, taking into account the dimension of the problem,
the propagator is
again a Dirac-delta with a sign change~:
\begin{equation}
    K(x_2,y_2,\tau|x_1,y_1, 0)=(-1)^N\delta\big(x_1-x_2,y_1-y_2\big). 
\end{equation}

It is worth mentioning that this calculation should turn out to be useful
to explain the Sagnac effect \cite{Sagnac}. This experiment, originally
proposed and performed with light and later been repeated with massive
particles, amounts to perform a two-slit type
interference experiment, when the whole apparatus is fixed on a
turntable. The clue is that the inertial force due to rotation
 behaves exactly as a fictious magnetic field, with twice the mass,
 $2m$, replacing the electric charge.

\section{The Semiclassical approximation}\label{szemi}

Returning to the general case, let us assume that $\cP$ has
a unique classical path,  $\bar{\gamma}$, 
and let us develop the action to second order
 \cite{LeSmi}~:
\begin{equation}
S(\gamma)=S(\bar{\gamma})+\delta_{\bar{\gamma}}(\eta)+
\half\delta^2S_{\bar{\gamma}}(\eta,\eta)+\dots
\label{hatkif}
\end{equation}
where the ``dots, $\dots$'', denote all higher-order
variations. 

Taking the semiclassical approximation
amounts of dropping all these terms.
According to Hamilton's Principle $\delta S_{\bar{\gamma}}=0$.
The semiclassical propagator is, hence,
\begin{equation}
K(\vec{x}_2,T|\vec{x}_1,0)=
\exp\left[\frac{i}{\hbar}S(\bar{\gamma})\right]\times
F(T),
\qquad 
F(T)=\int\!\exp\left\{\frac{i}{2\hbar}
\delta^2_{\bar{\gamma}}S(\eta,\eta)\right\}\cD\eta.
\label{SCpropag}
\end{equation}
cf. (\ref{oscpropag})-(\ref{F(T)}). 

The reduced propagator can be determined diagonalizing
the second variation. Let us assume, for simplicity, 
that the system is $1$-dimensional. The quadratic form
$\Lambda$ in the eigenvalue equation (\ref{sajert}) is
a self-adjoint Sturm-Liouville operator on the
space of all ``tangent vectors'' $\eta$.
The eigenvalues, $\lambda_n$, are therefore all real
and form a complete orthonormal system w.r.t.
the usual scalar product,
$
(\eta_{n},\eta_{m})=
T^{-1}\displaystyle\int\eta_{n}\eta_{m}dt=\delta_{nm}.
$ 
Expanding the variation $\eta$ as
$
\eta=\sum_{n}a_{n}\eta_{n}$ yields
$
\Big(\eta,\Lambda\eta\Big)=\sum_{k} a_{k}^2\lambda_{k}.
$
The reduced propagator is, therefore, once again
(\ref{sokintegral}), with the $\lambda_{k}$ denoting
the eigenvectors of $\Lambda$ in (\ref{sajert}).
Hence
\begin{equation}
F(T)=C\sqrt{\prod_k\frac{2i\pi\hbar}{\lambda_k}}\ .
\label{Jredprop}
\end{equation}
As explained before, the Jacobian is independent of the
dynamics, so that (\ref{Jredprop}) holds also for
the free factor,
\begin{equation}
F(T)^{free}=C\sqrt{\prod_k\frac{2i\pi\hbar}{\lambda_k^{free}}}\, ,
\label{szabJredprop}
\end{equation}
where 
 $F(T)^{free}$ and $\lambda_k^{free}$
 are the free propagator and the eigenvalues, respectively. Dividing
 (\ref{Jredprop}) by (\ref{szabJredprop}) $C$ drops out,
$$
F(T)=F(T)^{free}\times\sqrt{
{\displaystyle\prod_k\lambda_k^{free}}/{\displaystyle\prod_k\lambda_k\ \ \ \  }}\ \ .
$$
The semiclassical propagator is, therefore,
\begin{equation}
K=e^{iS(\bar{\gamma})}\left(\frac{m}{2\pi i\hbar T}\right)^{1/2}\times\sqrt{
\frac{\prod_k\lambda_k^{free}}{\,\prod_k\lambda_k\ \ \ \  }}\ ,
\label{szemiklprop}
\end{equation}
since, in $D$ dimensions,
$
F(T)^{free}=\left(\frac{m}{2\pi i\hbar T}\right)^{D/2},
$ 
cf. (\ref{szabadredprop}).

$\bullet$ For a $1D$ oscillator, we recover the previous
result~: by (\ref{tenyezok}), the quotient of
the products of the eignevalues under the root is
exactly the infinite product we determined using in
the Euler formula, since
 $\lambda_k^{free}=m\pi^2 k^2/T^2$. 

\goodbreak
\kikezd{The Van Vleck matrix}.

Our result can also be presented using the
{\it Van Vleck matrix} \cite{LeSmi}. Let us
indeed remember that, still assuming the
uniqueness of the classical path, $\bar{\gamma}$, between the
to given points $\vec{x}_{1}$-et $\vec{x}_{2}$ in time $T$, the
action can be viewed as function of the end points.
\begin{equation}
S(\vec{x}_{1},\vec{x}_{2})=S(\bar{\gamma})
\end{equation}
is  in fact Hamilton's Principal Function. The determinant of the
 ${D}\!\times\!{D}$ matrix
\begin{equation}
\left[\frac{\p^2 S}{\p \vec{x}_1\p \vec{x}_2}\right]=
    \left[\frac{\p^2 S}{\p x_{1}^i\p x_{2}^j}\right]
\end{equation}
 is called the {\it Van Vleck determinant}.
Then

\kikezd{Theorem} (\cite{LeSmi})~: In $D$ {\it dimensions,
the semiclassical propagator is
\begin{equation}
    K(\vec{x}_{2},T|\vec{x}_{1},0)
    =
    \left(\frac{1}{2\pi i\hbar}\right)^{1/2}
    \left|{\rm det}
    \frac{\p^2 S}{\p \vec{x}_1\p \vec{x}_2}\right|^{D/2}\!
    \exp\left[-iDN\frac{\pi}{2}\right]\times
    \exp\left[\frac{i}{\hbar}S(\bar{\gamma})\right].
\label{semiclassprop}
\end{equation}
where $\vec{x}_1=(x_1^i)$ and $\vec{x}_2=(x_2^j)$ are
the initial and final point, respectively.}

$\bullet$ For the $1D$ oscillator
\begin{equation}
\frac{\p^2 S}{\p x_{1}\p x_{2}}=-\frac{m\omega}{\sin\omega t},
\label{oscivV}
\end{equation}
so that (\ref{semiclassprop}) yields, once again,
the previous oscillator propagator formula.

$\bullet$  Let us consider a charged particle in a {\it constant
magnetic field}, and let us assume that $T\neq\tau$.
Hamilton's Principal function is now 
 (\ref{Bhatas}), and the absolute value of the Van Vleck 
determinant reads
\begin{equation}
\big|{\rm det}\,\frac{\p^2 S}{\p \vec{x}_{1}\p \vec{x}_{2}}\big|=
    m^2\omega^2\times{\rm det}\,\left(
    \begin{array}{cc}
    -\cot\omega T&1
    \\[6pt]
    -1&-\cot\omega T
    \end{array}\right)=
    \frac{m^2\omega^2}{\sin^2\omega T}.
    \label{BvVdet}
\end{equation}  
From (\ref{semiclassprop}) we get   (\ref{Bprop}) once again.

\section{Morse theory \cite{Morse}}

The methods of variational calculus allow us to
further deepen our understanding.
Let us chose a classical motion $\bar{\gamma}$ [in
$D$-dimensions], and let us focus our attention to the
second variation. As seen in  (\ref{Lambda})
$$
\half\delta^2S(\eta,\eta)=\int_{0}^T\!\big(\eta,\Lambda\eta\big)dt,
$$
where $\Lambda$  is the matrix of the second variation.
A {\it Jacobi field} along $\bar{\gamma}$ is a
$D$-dimensional vector  $\vec{\xi}(t)=(\xi_i(t))$
such that
\begin{equation}
\Lambda\,\vec{\xi}=0\qquad\hbox{i.e.}\qquad
\Lambda_{ij}\,\xi_j=0
\quad\forall\, i.
\label{Jacobifield}
\end{equation}

This second-order differential equation in $2D$ dimensions has
$2D$ independent solutions. Generally, none of them belongs to the tangent space of
$\cP$ at $\bar{\gamma}$, since $\vec{\xi}(0)$ and/or $\vec{\xi}(T)$
may not vanish. 
Two points  $p=\bar{\gamma}(t_1)$ and $q=\bar{\gamma}(t_1)$
of a curve $\bar{\gamma}$,
$0\leq t_1<t_2\leq T$ are called conjugate point, if there exists a Jacobi
field such that 
\begin{equation}
\vec{\xi}(t_1)=\vec{\xi}(t_2)=0,
\end{equation}
i.e., if it  belongs to the tangent space. The number of such independent
fields is called the multiplicity of the conjugate points $p$ and $q$.
 
The null space of the second variation is the set of all tangent
vectors $\eta$ 
for which
$
\half\delta^2S(\eta,\eta')=0
$
for all $\eta'$. The dimension, $\nu$, of the null space is thus the
number of independent Jacobi fields that also belong to the tangent space.
The null-space is, hence, non-trivial, if the starting and the ending
points of $\bar{\gamma}$, $p=x_1=\bar{\gamma}(0)$ and $q=x_2=\bar{\gamma}(T)$,
are conjugate.  $\nu$ is finite and it is readily seen that
$\nu<D$ \cite{Morse}.

It is easy to see that $2D$ independent Jacobi fields can be
constructed using classical paths. Let us indeed consider a one-paramater
family of classical paths [i.e. solutions of the  Euler-Lagrange equations]
$\beta(u),\ 0\leq u\leq 1$, such that
$
\beta(0)=\bar{\gamma}.
$
These paths are not required to have the same end points as $\bar{\gamma}$.
Then it is easy to see that 
\begin{equation}
\vec{\xi}=\frac{\p\beta}{\p u}\bigg|_{u=0}
\end{equation}
is a Jacobi field along $\bar{\gamma}$. The general solution
of the Euler-Lagrange equations depends on $2D$ parameters.
Deriving w.r.t. these parameters yields  $2D$ independent
Jacobi fields.

Let us now define the {\it Morse index} of $\bar{\gamma}$ as the
maximal dimension,  $\mu$, of those subspaces of the tangent space
upon which  the  restriction of the second variation
is negative definite. 
\begin{equation}
\delta^2S_{\gamma}(\eta,\eta)<0.
\end{equation}
Then we have~:

\kikezd{Theorem} (Morse) \cite{Morse}~: {\it 
Morse index is the number of those points
of $\bar{\gamma}$,  each counted with its multiplicity,
 which are conjugate to  $x_1=\bar{\gamma}(0)$.}
\vskip2mm

The essence of the proof is to show that the number of negative
eigenvalues is precisely the Morse index $\mu=\mu(T)$.
This is proved by showing that the Morse index, $\mu(t)$, calculated on the
segment from  $t=0$ to $t$ of the curve is a monotonic function of $t$
which, for sufficiently small $t$, is continuous from the left,
$
\mu(t-\epsilon)=\mu(t)
$
for $\epsilon>0$ sufficiently small. From the right, instead,
\begin{equation}
\mu(t+\epsilon)=\mu(t)+\nu,
\label{mujump}
\end{equation}
where $\nu$ is, as before, the dimension of the
null space of the second variation. The $\mu(t)$ as a function of
time is hence, at first,  constant, and then, upon crossing a
conjugate point, it jumps by $\nu$.

The Morse index is finite, so that our curve contains a finite
number of points conjugate to the starting point.

Let us again illustrate the general theory on example.

$\bullet$ For the $1D$ oscillator and for
 $\Lambda$ in  (\ref{oscLambda}),
the solutions of the Jacobi equation  (\ref{Jacobifield}) are
$
a\sin\omega t+b\cos\omega t,
$
which vanish at $t=0$  and at $t=T$ if $b=0$  and
$T\omega=N\pi$. A point $q$ of the classical trajectory
is conjugate to $x_1=\bar{\gamma}(0)$ if
$$
q=q_k=\bar{\gamma}(k\frac{\tau}{2}),
\qquad
k=1,\dots
$$
Consistently with our general theorem, the null space of $\delta^2S$
can be also obtained by considering the classical motion which starts from
$x_1$ with velocity  $v$,
\begin{equation} 
\frac{v}{\omega}\sin\omega t+ x_1\cos\omega t\quad\Rightarrow\quad
\xi(t)=\frac{\p x(t)}{\p v}=\frac{1}{\omega}\sin\omega t.
\end{equation} 
For $t=N\tau/2$, we get manifestly a null-mode of the second
variation.

For a $D$-dimensional oscillator, each conjugate point contributes
$D$ null-modes. When $T$ crosses $2N/\tau$ the dimension of the null space,
i.e., the Morse index jumps by $D$.

Let us mention that, on compact manifolds,  the Morse index is related
to the topology of the infinite dimensional manifold of
paths \cite{Morse}.

\section{Conclusion}

Is the phase observable~?
The naive answer to this question is negative: when the wave function is
multiplied by any phase factor, the absolute value of the amplitude,
and hence its square, the probability, are unchanged.
This is only true for one wave function, though. If the system is
decomposed into parts and the partial waves are multiplied by 
different phase factors, recombination will yield an interference.

Similar phase jumps are found in molecular \cite{molek}, nuclear \cite{nukl}, 
and heavy ion \cite{heavyion} collisions.

After the original ideas reviewed here have been put forward,
several  developments took place. See,  e.g. \cite{Tsutsui,noncons}. 
 Related questions are also discussed in some textbooks \cite{books}.
 
It is worth calling the attention to that all path integrals studied in this
 paper can  be calculated \cite{Inomata} by transforming the system
 into the free form using a ``non-relativistic conformal transformation''
 \cite{Niederer,DHconf}. For the harmonic oscillator, for example,
 every half-period can be conformally related to a full free motion.
 Then the Maslov correction can be recovered
 by fitting together the wave function at half-periods
 \cite{DHconf}. This latter paper also provides a complete catalog of 
 all possibilities.



\end{document}